\documentclass[twocolumn,showpacs,preprintnumbers,amssymb,nofootinbib]{revtex4}

\usepackage{graphicx,epsf, epsfig, amssymb}% Include figure files
\usepackage{bm}
\usepackage{longtable}
\usepackage{color}
\usepackage{amsfonts,amsmath,amssymb,mathrsfs}

\def\be{\begin{equation}}
\def\ee{\end{equation}}
\def\beq{\begin{eqnarray}}
\def\eeq{\end{eqnarray}}

\begin{document}

\title{Test bodies and naked singularities: is the self-force the cosmic censor?}


\author{
Enrico Barausse$^{1}$, %\footnote{Electronic address: barausse@umd.edu}$,
Vitor Cardoso$^{2,3}$, %\footnote{Electronic address: vitor.cardoso@ist.utl.pt}$, 
Gaurav Khanna$^{4,5}$,% \footnote{Electronic address: khanna@umassd.edu}$
}

\affiliation{${^1}$ Department of Physics, University of Maryland, College
Park, MD 20742, USA}

\affiliation{${^2}$ CENTRA, Departamento de F\'{\i}sica, %Instituto Superior T\'ecnico, 
IST/UTL,
%Universidade T\'ecnica de Lisboa - UTL,
Av.~Rovisco Pais 1, 1049 Lisboa, Portugal}

\affiliation{${^3}$ Department of Physics and Astronomy, The University of
Mississippi, University, MS 38677, USA}

\affiliation{${^4}$ Physics Department, University of Massachusetts
  Dartmouth, North Dartmouth, MA 02747, USA}

\affiliation{${^5}$ Albert-Einstein-Institut, Max-Planck-Institut f¨ur Gravitationsphysik, Hannover, Germany}

\begin{abstract}
Jacobson and Sotiriou showed that rotating black holes could be
spun-up past the extremal limit by the capture of non-spinning test bodies, if one neglects radiative and self-force effects.
This would represent a violation of the Cosmic Censorship Conjecture in
four-dimensional, asymptotically flat spacetimes.  
We show that for \textit{some} of the trajectories giving rise to naked singularities, radiative effects \textit{can} be neglected. 
However, for these orbits the conservative self-force is important, and seems to have the right sign to
prevent the formation of naked singularities.
\end{abstract} 

\maketitle

%%%%%%%%%%%%%%%%%%%%%%%%%%%%%%%%%%%%%%%%%%%%%%%%%%%%%%%%%%%%%%%%%%%%%%%%%%%%%%%%%%%%%%%%%%%%%%%%%%%%
% \section{Introduction}
%%%%%%%%%%%%%%%%%%%%%%%%%%%%%%%%%%%%%%%%%%%%%%%%%%%%%%%%%%%%%%%%%%%%%%%%%%%%%%%%%%%%%%%%%%%%%%%%%%%%
The most general stationary vacuum black-hole (BH) solution of Einstein's equations in a
four-dimensional, asymptotically flat spacetime is the Kerr geometry~\cite{Kerr:1963ud}, characterized only by its
mass $M$ and angular momentum $J$. Solutions spinning below the Kerr
bound $cJ/GM^2 \leq 1$ possess an event horizon and are known as Kerr
BHs. Solutions spinning faster
than the Kerr bound describe a ``naked singularity'', where classical
General Relativity breaks down and (unknown) quantum gravity effects
take over. It was hypothesized by Penrose that classical General
Relativity encodes in its equations a mechanism to save it from the
breakdown of predictability. This is known as the Cosmic Censorship
Conjecture (CCC)~\cite{Wald:1997wa}, which asserts that every singularity
is cloaked behind an event horizon, from which no information can
escape.

There is no proof of the CCC. Indeed there are a few known
counter-examples, but these require either extreme fine-tuning in the
initial conditions or unphysical equations of state~\cite{Wald:1997wa}, or are staged in
higher-dimensional spacetimes~\cite{Lehner:2010pn}. Moreover, all existing evidence indicates
that Kerr BHs are perturbatively stable~\cite{Whiting:1988vc}, while Kerr
solutions with $cJ/GM^2 > 1$ are unstable~\cite{unstable_naked}.
Thus, naked singularities cannot form from BH instabilities.

Because naked singularities appear when $cJ/GM^2>1$, it is
conceivably possible to form them by throwing matter with sufficiently large angular momentum into a BH. 
With numerical-relativity simulations, Ref.~\cite{Sperhake:2009jz} 
found no evidence of formation of naked singularities in a high-energy collision between two comparable-mass BHs:
either the full nonlinear equations make the system
radiate enough angular momentum to form a single BH, or the BHs simply scatter. 
The case of a test-particle plunging into an \textit{extremal} Kerr BH was studied by
Wald~\cite{wald}, who showed that naked singularities can never be produced, because particles carrying
dangerously large angular momentum are just not captured. 

Recently, Jacobson and Sotiriou (JS)~\cite{Jacobson:2009kt}
(building on Refs.~\cite{previous}) have
shown that if one considers an \textit{almost} extremal BH, non-spinning particles carrying enough angular momentum
to create naked singularities \textit{are} allowed to be captured.\footnote{JS consider also the case of spinning particles, but in this
letter we will focus on the non-spinning case.} As acknowledged by JS, however,
their analysis neglects the  conservative and dissipative self-force (SF), and both effects may be important~\cite{hod}. In this letter we will show 
that the dissipative SF (equivalent to radiation reaction, \textit{i.e.} 
the energy and angular momentum losses through gravitational waves)
can prevent the formation of naked singularities only for some of JS's orbits. However, we will show that for \textit{all}
these orbits the conservative SF is comparable to the terms giving rise to naked singularities,
and should therefore be taken into account. Hereafter we set $G=c=M=1$.

JS considered a BH with spin
$a\equiv J/M^2=1-2 \epsilon^2$, with $\epsilon \ll 1$, and a non-spinning test-particle with energy $E$, angular momentum $L$ and mass $m$.
Neglecting the dissipative and conservative SF, the particle moves on a geodesic, and
JS identified a class of equatorial geodesic orbits such that
\textit{(i)} the particle falls into the BH, which implies an upper limit on the angular momentum, $L<L_{\max}$, and
\textit{(ii)} the BH is spun up past the
extremal limit and  destroyed, which implies a lower limit on the angular momentum, $L>L_{\min}$. Therefore
\beq
L_{\min}=2 \epsilon^2 + 2 E + E^2<L< L_{\max}= (2 + 4 \epsilon) E\,.\label{limitsJ}
\eeq
Imposing $L_{\max}>L_{\min}$ then yields  
\beq
E_{\rm min}= (2 - \sqrt{2})\epsilon<E< E_{\max}= (2 + \sqrt{2})\epsilon\,.\label{limitsE}
\eeq
Finally, JS checked that these intervals contain both {\it bound} orbits (\textit{i.e.} orbits that start with zero radial velocity at finite radius) and {\it unbound} orbits (\textit{i.e.} orbits that start from infinity).
Parameterizing the above interval as
\begin{align}
&E=E_{\min}+x (E_{\max}-E_{\min})= E_{\min}+ 2 x \sqrt{2} \epsilon\label{Ex}\\
&L=L_{\min}+y (L_{\max}-L_{\min})=L_{\min} + 8 y \epsilon^2 (1-x) x\label{Ly}
\end{align}
with $0<x<1,\,0<y<1$, the final spin is
\be
a^{JS}_f=\frac{a+L}{(1+E)^2}= 1+8 \epsilon^2 (1-x) x y+{\cal O}(\epsilon^3)>1\,,\label{af}
\ee
and the spin-up is due to the terms quadratic in $\epsilon$.

Let us first investigate how radiation reaction 
affects JS's analysis. Taking radiation losses $E_{\rm rad}$ and $L_{\rm rad}$ into account, Eq.~\eqref{af} becomes
\be
a_f=
1+8 \epsilon^2 (1-x) x y+2E_{\rm rad}-L_{\rm rad}+{\cal O}(\epsilon^3)\,.
\label{radcorrections}
\ee
Let us focus on {\it unbound} geodesics,\footnote{As we mentioned, JS also considered bound orbits, falling 
into the BH from a Boyer-Lindquist radius $r=r_{\rm hor}+{\cal O}(\epsilon)$ 
($r_{\rm hor}$ being the horizon's radius). However, these orbits pose a problem,
as we will show later, 
because the distance to the horizon is comparable to the particle's
\textit{minimum} attainable size $\max(E,m)\gtrsim \epsilon$, so finite-size effects should be taken into account.}
and following JS assume $E/m\gg1$ and $L/m\gg1$ (null orbits). These orbits are 
characterized by the impact parameter $b=L/E$ alone.
From Eqs.~\eqref{limitsJ} and~\eqref{limitsE}, 
JS's orbits have $L=bE$, with 
$b=2 + 4 \epsilon [1 - 2 x (x - 1) (y - 1)/(2 + \sqrt{2} (2 x - 1))]$.
Varying $x$ and $y$ between 0 and 1, one obtains 
$b=2+\delta\epsilon$, with $2 \sqrt{2}<\delta<4$.
However, because $b_{\rm ph}=2+2 \sqrt{3}\epsilon+{\cal O} (\epsilon^2)$ is the impact parameter
of the circular photon orbit (``light-ring''),  only orbits with $2 \sqrt{2}<\delta< 2 \sqrt{3}$ are unbound.
When $\delta\approx2 \sqrt{3}$, these orbits are expected to circle many times 
around the light-ring, so radiation reaction could prevent the formation of naked singularities 
or at least invalidate
JS's analysis. In fact, for $\delta$ {\it arbitrarily} close to $2 \sqrt{3}$, 
the particles would orbit around the light-ring an arbitrarily large number of times, 
and gravitational-wave emission \textit{must} be important~\cite{Berti:2009bk}. 
We will show, however, that this is \textit{not} true for all of JS's orbits.

Considering the geodesic equations for null equatorial orbits with impact parameter
$b=b_{\rm ph} (1-k)$, with $k\ll\epsilon \ll1$, one finds that
the radial potential -- defined as $V_r(r)\equiv (dr/d\lambda)^2$ with $\lambda$ an affine parameter  -- has a minimum
at $r=r_{\min}=r_{\rm ph}+{\cal O}(k)$, near which
\be
\frac{d\phi}{dr}\approx \left(\frac83+\frac{\sqrt{3}}{2 \epsilon}\right)\left[\frac{8}{\sqrt{3}} k\epsilon+3 (r-r_{\min})^2\right]^{-1/2}\,.
\ee
Integrating from $r_{\min}-\Delta r_2$ to $r_{\min}+\Delta r_1$, with $\Delta r_{1,2}\gg k \epsilon$, the number of cycles near the minimum is
\begin{equation}
\label{cycles}
N_{\rm cycles} \approx \int^{r_{\min}+\Delta r_1}_{r_{\min}-\Delta r_2} \frac{d\phi}{dr}  \frac{dr}{2 \pi}  =[A+B \log{(k \epsilon)}]\left(\frac83+\frac{\sqrt{3}}{2 \epsilon}\right)
\end{equation}
$A$ and $B$ being constants depending on the integration interval. 
Fixing $\epsilon$, and thus the BH spin, we can see that $N_{\rm cycles}$ depends on $\log k$, 
and diverges when $k\to0$~\cite{Berti:2009bk}.

Because the fluxes are proportional to $N_{\rm cycles}$, we have
\be\label{totalfluxes}
E_{\rm rad}= \Delta E(\epsilon)\times N_{\rm cycles}\,,\quad
L_{\rm rad}= \Delta L(\epsilon)\times N_{\rm cycles}\,,
\ee
where $\Delta E$ and $\Delta L$ are the fluxes in a single orbit.
From a frequency-domain analysis~\cite{Teukolsky:1972my}, $\Delta E/\Delta L$ must equal the light-ring frequency, $\Omega_{\rm ph}\approx1/2-(\sqrt{3}/2)\epsilon$, hence
\beq
\Delta E(\epsilon)&=&E_1(\epsilon) (1+e_2 \epsilon)\,,\label{scalingE}\\
\Delta L(\epsilon)&=&2 E_1(\epsilon) [1+(\sqrt{3} + e_2) \epsilon]\label{scalingL}\,.
\eeq
Here $E_1(\epsilon)$ is the energy flux for a single orbit  at leading order in $\epsilon$, and $e_2$ is an undetermined coefficient.
Semi-quantitative arguments by Chrzanowski \cite{Chrzanowski:1976jy} 
and more rigorous analytical calculations by Chrzanowski and Misner \cite{Chrzanowski:1974nr} show that,
\be\label{E1scaling}
E_1\sim (r-r_H) E^2\sim \epsilon E^2 \sim \epsilon^3\,.
\ee
(later we will discuss an additional proof of this scaling).
At leading order in $\epsilon$, this results in
\be
E_{\rm rad}= \Delta E(\epsilon) \times N_{\rm cycles}\sim \log{(k \epsilon)} \epsilon^2\,.
\ee
This scaling still depends on $k$, but the dependence is logarithmic, so unless $k$ is really small
$E_{\rm rad} \sim \log{(\epsilon)} \epsilon^2$. Although
terms of order $\epsilon^2\log\epsilon$ seem to dominate Eq.~(\ref{radcorrections}),
because of Eqs.~\eqref{totalfluxes},~\eqref{scalingE} and \eqref{scalingL} one has 
$L_{\rm rad}-2E_{\rm rad}= 2 \sqrt{3}\epsilon E_1(\epsilon) N_{\rm cycles}
\sim \epsilon^3\log\epsilon$. Therefore JS's analysis is valid for these trajectories.
However, if $k\lesssim \exp(-1/\epsilon)$, $E_{\rm rad}\sim \epsilon$ and $L_{\rm rad}-2E_{\rm rad}\sim \epsilon^2$, 
and JS's analysis is not valid because radiative effects cannot be neglected. 

To test the above picture we used a time-domain code~\cite{TDEMRICode} 
solving the inhomogeneous Teukolsky equation~\cite{Teukolsky:1972my} that describes 
the gravitational perturbations of Kerr BHs in the context of extreme mass-ratio binaries (EMRBs). 
This code has been successfully used in many scenarios, including an extensive study of recoil velocities from EMRBs~\cite{RecoilEMRI}. 
% is a finite-difference, hyperbolic PDE solver using the Lax-Wendroff explicit numerical 
% evolution scheme. It 
Because, for almost extremal BHs and in Boyer-Lindquist coordinates, the particle's orbit, the light-ring and the 
horizon are extremely close, we modeled the test-particle to have a fixed width in the 
``tortoise'' coordinate $r^*$ as opposed to $r$~\cite{TDEMRICodeMIT12},
and checked that our results are independent of the particle's width when that is sufficiently small. 
(More details on these tests will be presented in a follow-up paper.) 

We consider BHs with $a=0.99$, $0.992$, $0.994$, $0.996$, $0.998$ and $0.999$ and geodesics having $E=(E_{\max}+E_{\min})/2=2\epsilon$, $L=b_{\rm ph}E (1-k)$ with $k=10^{-5}$, and $m=0.001 \ll E$.
Using these geodesics and integrating their cycles from $r=1.05 r_{\rm ph}$ to $r=(r_{\rm ph}+r_{\rm hor})/2$ ($r_{\rm ph}$ being the light-ring radius), 
we get $A\approx 0.3294$, $B\approx-0.01941$
for the coefficients in Eq.~\eqref{cycles}. Assuming $E_1= e_1 \epsilon^n$, we fit the energy and angular-momentum 
fluxes at infinity with 
Eqs.~\eqref{totalfluxes}-\eqref{scalingL}, obtaining $n\approx 2.91$. Because this is very close to the theoretical value $n=3$, we assume 
$n=3$ and fit the data with only two free parameters, $e_1$ and $e_2$,
obtaining $e_1=136.97$ and $e_2=-4.423$. With these values, Eqs.~\eqref{totalfluxes}-\eqref{scalingL} 
reproduce the numerical data to within $1-3\%$ for $a<0.999$, which is comparable to  the data
accuracy. For $a=0.999$, however, the fluxes predicted by Eqs.~\eqref{totalfluxes}-\eqref{scalingL} are about $12\%$ larger than
the numerical ones. To investigate this issue, we ran an additional
simulation for $a=0.9998$, which seems to confirm that Eqs.~\eqref{totalfluxes}-\eqref{scalingL} overpredict the fluxes for very high spins.
At this stage it is not clear whether this is a numerical problem (simulations are very challenging for 
$a\approx1$) or whether this is due to the simplified analytical derivation of 
Eqs.~\eqref{totalfluxes}-\eqref{scalingL}. We will investigate this issue in the follow-up paper, but 
because the numerical fluxes are \textit{smaller} than expected, it only reinforces our conclusion
that there are orbits giving rise to naked singularities \textit{even} 
when radiation reaction is taken into account.

Since $L_{\rm rad}-2E_{\rm rad}= 2 \sqrt{3}\epsilon E_1(\epsilon) N_{\rm cycles}
\sim \epsilon^3\log(k\epsilon)>0$, Eq.~\eqref{radcorrections} predicts that radiation reaction will decrease the final spin $a_f$.
Using the above values for $A,\,B$ and $e_1$, 
and $x=0.5$ and $y\approx  2 \sqrt{3}-3+4 \epsilon/3$ corresponding to our geodesics, 
Eq.~\eqref{radcorrections} predicts $a_f<1$ for $\epsilon\gtrsim\epsilon_{\rm crit}\sim0.003$. However, for sufficiently 
large spins, the term $L_{\rm rad}-2E_{\rm rad}\sim \epsilon^3\log(\epsilon)$ is subdominant
and $a_f>1$. Numerical results confirm this expectation: in Table~\ref{tab:jfin}, 
we show the BH spin $a_f$ after absorbing
the particle, taking into account radiation reaction. As can be seen, $a_f>1$ already for $a=0.9998$, corresponding to $\epsilon=0.01>\epsilon_{\rm crit}$. This is because, as already mentioned,
Eqs.~\eqref{totalfluxes}-\eqref{scalingL} overpredict the fluxes for $a\gtrsim 0.999$.
\begin{table}
\caption{Initial and final BH spin after absorbing a particle with energy $E=\sqrt{2(1-a)}$ and angular momentum $L=b_{\rm ph}E (1-10^{-5})$, neglecting conservative SF effects, but not radiation reaction. We also show the final spin without radiation reaction ($a_f^{JS}$) predicted by JS.}
\begin{tabular}{rccccccc}\hline
$a$             & 0.99   & 0.992   & 0.994  & 0.996  &  0.998 & 0.999 & 0.9998 \\
$a_f$           & 0.882  & 0.928    & 0.961 & 0.984 & 0.997 & 0.9996 & 1.00006 \\
$a_f^{JS}$       & 1.0043 & 1.0035    & 1.0026 & 1.0018 & 1.0009 & 1.00045 & 1.00009 \\
\hline \hline
\end{tabular}
\label{tab:jfin}
\end{table}

Moreover, even the result that $a_{f}<1$ for $\epsilon\gtrsim\epsilon_{\rm crit}$ is questionable. Indeed, the 
fluxes down the horizon might destroy the BH before the particle is captured, while our code only calculates the fluxes at infinity.
This is sufficient for our purposes because we used the code only  to 
test the scaling~\eqref{totalfluxes}-\eqref{scalingL}, which is expected to hold
\textit{both} for the fluxes at infinity and down the horizon, since its derivation is generic. 
Once validated, that scaling implies that for 
sufficiently large spins both fluxes are smaller than the terms giving rise to naked
singularities [\textit{i.e.} the quadratic terms in Eq.~\eqref{radcorrections}]. 
For $\epsilon\gtrsim\epsilon_{\rm crit}$, instead,  the fluxes at infinity
decrease the final spin to $a_{\rm fin} <1$. However, 
in such a situation also the fluxes down the horizon, $L_{\rm rad,in}$ and $E_{\rm rad,in}$, are expected to
be important (because the fluxes are produced when the particle sits at the light-ring, which roughly corresponds to the maximum of
the effective potential for gravitational waves), and could destroy the horizon \textit{before} the particle is
captured. In fact, the spin change is $\Delta a=L_{\rm rad,in}-2E_{\rm rad,in}=E_{\rm rad,in} (1/\Omega_{\rm ph}-2)\approx 2 \sqrt{3}\epsilon E_{\rm rad,in}$, because $E_{\rm rad,in}/ L_{\rm rad,in}=\Omega_{\rm ph}\approx1/2-(\sqrt{3}/2)\epsilon$. Since $\Omega_{\rm ph}$ is larger than the horizon's frequency $\Omega_{\rm hor}\approx 1/2-\epsilon$, radiative emission is non-superradiant and $E_{\rm rad,in}>0$, hence $\Delta a>0$. Thus, the ingoing fluxes \textit{increase} $a_f$.

So far we have shown that radiation reaction cannot prevent the formation of naked singularities, 
unless the impact parameter is extremely close to
the light-ring's impact parameter $b_{\rm ph}$. We will now show, however, that for \textit{all} of JS's orbits  
the conservative SF is as important as the terms giving rise to naked singularities.

Let us consider a BH\footnote{This discussion is completely general because the motion
of a BH is the same as that of a particle with mass $m$, at leading and next-to-leading order in $R_g/{\cal L}$~\cite{MST}.} with gravitational radius $R_g = 2 G m/c^2$ in a 
curved background with curvature radius ${\cal L}\gg R_g$. 
The rigorous way of studying the motion of this BH
is to set up a proper initial value formulation, but a  reasonable
alternative for practical purposes is to use a matched asymptotic expansion~\cite{MST}.
Near the BH (\textit{i.e.} for $r< r_i$, $r_i$ being a radius $\ll {\cal L}$), the metric is $g_{\rm internal} = g_{\rm BH} +H_1(r/{\cal L})+H_2(r/{\cal L})^2+...$,
where $g_{\rm BH}$ is the metric of an isolated BH and $H_1(r/{\cal L})$, $H_2(r/{\cal L})^2$ are corrections due to the ``external'' background.
Far from the BH (\textit{i.e.} for $r>r_e$, $r_e$ being a radius $\gg R_g$), the metric is instead $g_{\rm external} = 
g_{\rm background} +h_1 (R_g/{\cal L})+h_2 (R_g/{\cal L})^2+...$, 
\textit{i.e.} the background metric plus perturbations due to the BH's presence. 
Because $R_g \ll {\cal L}$, there exists a region $r_e<r<r_i$ where both pictures are valid and the two metrics can be matched. Doing so, one finds 
that the BH equations of motion are~\cite{MST}
\begin{equation} \label{sf}
u^\mu \nabla_\mu u^\nu = f^\nu_{\rm cons}+f^\nu_{\rm diss}+{\cal O}(R_g/{\cal L})^2\,,
\end{equation} 
where $\nabla$ is the connection of the background spacetime. The terms $f^\nu_{\rm cons}$ 
and $f^\nu_{\rm diss}$ are ${\cal O}(R_g/{\cal L})$, and are known as the conservative and dissipative SF. Remarkably, it turns
out that Eq.~\eqref{sf} is the geodesic equation of a particle in a ``perturbed'' metric $\tilde{g}=g+h^R$, where $h^R$ is a smooth tensor field
of order ${\cal O}(R_g/{\cal L})$:
\begin{equation}  \label{sf2}
\tilde{u}^\mu \tilde{\nabla}_\mu \tilde{u}^\nu = 0\,
\end{equation} 
(the connection $\tilde{\nabla}$ and the 4-velocity $\tilde{u}^\mu$ being defined with respect to the ``perturbed'' metric $\tilde{g}=g+h^R$).

The dissipative SF amounts to the energy and angular-momentum fluxes 
considered earlier. Taking for instance the energy loss, Eq.~\eqref{sf} and $E=-p_t$ give 
$dE/d\tau= - m f^{\rm diss}_t ={\cal O}(R_g/{\cal L})^2$.
Assuming now that the background spacetime is a BH with mass $M\sim {\cal L} \gg R_g$, and specializing to orbits near the horizon, 
one has $dt/d\tau \sim r_H/(r-r_H)$, which gives $dE/dt\sim (r-r_H) {\cal O}(R_g/{\cal L})^2$.
Comparison of this scaling with our numerically validated scaling~\eqref{E1scaling} shows that for a BH with $E\gg m$
the size entering the matched asymptotic expansion above (the ``physical'' size)  is $R_g = 2 G E/c^2$ and \textit{not} $R_g = 2 G m/c^2$. 
This is no surprise, as the physical size associated with an
ultrarelativistic BH is dictated by its energy and not by its mass, because in General Relativity energy gravitates.
Remarkably, however, we were able to \textit{test} this fact with the numerical results presented earlier.
Further evidence comes from boosting the Schwarzschild line-element to the speed of light, keeping the total energy fixed. 
One gets the Aichelburg-Sexl metric, which depends on the total energy $E$ and not on the rest-mass~\cite{Aichelburg:1970dh}: this boosted BH
absorbs particles within a distance $\sim E$ from it. %, so measurements cannot discriminate lengths smaller than $\sim E$.

Because a BH's size is  determined by  $\max(E,m)\gtrsim \epsilon$, 
the conservative SF affects 
JS's analysis. This is easier to see from Eq.~\eqref{sf2} 
(although the same result can be obtained from Eq.~\eqref{sf}: see Ref.~\cite{barack_sago}): 
because the metric ``perturbation'' $h^R$ is ${\cal O}(R_g/{\cal L})={\cal O}(\epsilon)$,
the effective potential for the radial motion differs from the ``geodetic'' one by   
${\cal O}(R_g/{\cal L})={\cal O}(\epsilon)$~\cite{barack_sago,Sago:2008id}.
Therefore,  $b_{\rm ph}$ changes by $\delta b_{\rm ph} ={\cal O}(R_g/{\cal L})={\cal O}(\epsilon)$. 
Because JS's orbits have $b_{\rm ph}-b={\cal O}(R_g/{\cal L})={\cal O}(\epsilon)$, 
the conservative SF may prevent them from plunging into the horizon. This effect is intuitive: if the particle's 
size is $\sim \epsilon$, finite-size effects are important for impact parameters $b=b_{\rm ph}+{\cal O}(\epsilon)$.

A calculation of $\delta b_{\rm ph}$ is not doable with present technology~\cite{SFcodes}, 
but we can estimate its sign. Because of frame-dragging, for $a=1-2 \epsilon^2$ one has $b_{\rm ph}= 1/\Omega_{\rm ph}+ {\cal O}(\epsilon)^2$.
While the SF effect on $\Omega_{\rm ph}$ has not been calculated yet, Ref.~\cite{barack_sago} calculated the
Innermost Stable Circular Orbit (ISCO)-frequency shift for $a=0$, and showed that the conservative SF increases $\Omega_{\rm ISCO}$. It therefore seems
plausible that $\Omega_{\rm ph}$ should follow the same behavior.
While approximate methods for calculating the conservative SF in Kerr spacetimes exist~\cite{eob,favata}, they have problems for 
large spins, and the definitive answer to whether $b_{\rm ph}$ increases or decreases for $a\approx1$ will only be available when a
rigorous SF calculation~\cite{SFcodes} is performed. However, assuming that the $a=0$ behaviour of Ref.~\cite{barack_sago}
holds also for $a\approx1$, one obtains that $\Omega_{\rm ph}$ increases due to the SF, and therefore $b_{\rm ph}$ should decrease,
possibly preventing the capture of the particles with dangerously large $L$ and the formation of naked singularities.

  In conclusion, we have shown that radiation reaction effects can prevent the formation of naked singularities
   only for \textit{some} of the orbits for non-spinning particles around almost extremal Kerr BHs identified by JS.
    However, for \textit{all} orbits capable of producing a naked singularities, the conservative SF is 
     non-negligible and seems to have the right sign to prevent the particles from being captured, thus saving the Cosmic Censorship Conjecture.

 \noindent\textit{Acknowledgements:} We thank S. A. Hughes for enlightening discussions and for testing part of our analysis
with his frequency-domain Teukolsky code, and L. Barack, E. Berti, L. Gualtieri, T. Jacobson, S. Liberati, F. Pretorius, U. Sperhake and T. P. Sotiriou for reading 
this manuscript and providing useful comments. E.B. and G.K. acknowledge support from NSF Grants PHY-0903631 and PHY-0902026, respectively. 
This work was supported by {\it DyBHo--256667} ERC Starting, NSF PHY-090003 and FCT - Portugal through PTDC projects FIS/098025/2008, FIS/098032/2008, CTE-AST/098034/2008, and CERN/FP/109290/2009. 


\begin{thebibliography}{99}

\bibitem{Kerr:1963ud}
  R.~P.~Kerr,
  Phys.\ Rev.\ Lett.\  {\bf 11}, 237 (1963).

\bibitem{Wald:1997wa}
  R.~M.~Wald,
  arXiv:gr-qc/9710068.

\bibitem{Lehner:2010pn}
  L.~Lehner and F.~Pretorius,
Phys.\ Rev.\ Lett.\  {\bf 105}, 101102 (2010)
%arXiv:1006.5960 [hep-th].
  
\bibitem{Whiting:1988vc}
  B.~F.~Whiting,
  J.\ Math.\ Phys.\  {\bf 30}, 1301 (1989);
  E.~Berti, V.~Cardoso and A.~O.~Starinets,
  Class.\ Quant.\ Grav.\  {\bf 26}, 163001 (2009).

 \bibitem{unstable_naked} P.~Pani  {\it et al},
   Phys. Rev. D {\bf 82}, 044009 (2010);   G.~Dotti  {\it et al},
   Class.\ Quant.\ Grav.\  {\bf 25}, 245012 (2008);  V.~Cardoso  {\it et al},
   Class.\ Quant.\ Grav.\  {\bf 25}, 195010 (2008).

\bibitem{Sperhake:2009jz} U.~Sperhake {\it et al}, % V.~Cardoso, F.~Pretorius, E.~Berti, T.~Hinderer and N.~Yunes,
  Phys.\ Rev.\ Lett.\  {\bf 103}, 131102 (2009);
  M.~Shibata, H.~Okawa and T.~Yamamoto,
  Phys.\ Rev.\  D {\bf 78}, 101501 (2008).

\bibitem{wald} R.~M.~Wald, 
Ann. Phys. {\bf 82}, 548 (1974)%;  M.~Bouhmadi-Lopez {\it et al},
%   Phys.\ Rev.\  D {\bf 81}, 084051 (2010).

\bibitem{Jacobson:2009kt}
  T.~Jacobson and T.~P.~Sotiriou,
  Phys.\ Rev.\ Lett.\  {\bf 103}, 141101 (2009).

\bibitem{previous}
  V.~E.~Hubeny,
  Phys.\ Rev.\  D {\bf 59}, 064013 (1999); 
  S.~Hod,
  Phys.\ Rev.\  D {\bf 66}, 024016 (2002).

\bibitem{hod} S.~Hod, Phys. Rev. Lett. 100, 121101 (2008) 

\bibitem{Berti:2009bk} E.~Berti {\it et al}, %V.~Cardoso, L.~Gualtieri, F.~Pretorius and U.~Sperhake,
  Phys.\ Rev.\ Lett.\  {\bf 103}, 239001 (2009); E.~Berti {\it et al}, % V.~Cardoso, T.~Hinderer, M.~Lemos, F.~Pretorius, U.~Sperhake and N.~Yunes,
  Phys.\ Rev.\  D {\bf 81}, 104048 (2010).

\bibitem{Teukolsky:1972my}
  S.~A.~Teukolsky,
  Phys.\ Rev.\ Lett.\  {\bf 29}, 1114 (1972).

\bibitem{Chrzanowski:1976jy}
  P.~L.~Chrzanowski,
  Phys.\ Rev.\  D {\bf 13}, 806 (1976).

\bibitem{Chrzanowski:1974nr}
  P.~L.~Chrzanowski and C.~W.~Misner,
  Phys.\ Rev.\  D {\bf 10}, 1701 (1974).

\bibitem{TDEMRICode}
  L.~M.~Burko and G.~Khanna,
  Europhys.\ Lett.\ {\bf 78}, 60005 (2007).


\bibitem{RecoilEMRI}
  P.~A.~Sundararajan, G.~Khanna and S.~A.~Hughes,
  Phys.\ Rev.\ D {\bf 81}, 104009 (2010).

\bibitem{TDEMRICodeMIT12}
  P.~A.~Sundararajan, G.~Khanna and S.~A.~Hughes,
  Phys.\ Rev.\ D {\bf 76}, 104005 (2007); P.~A.~Sundararajan  {\it et al},
Phys.\ Rev.\ D {\bf 78}, 024022 (2008).

\bibitem{MST}
  Y.~Mino, M.~Sasaki and T.~Tanaka,
  Phys.\ Rev.\  D {\bf 55}, 3457 (1997); S.~E.~Gralla and R.~M.~Wald, arXiv:0907.0414;  R.~M.~Wald, arXiv:0907.0412 

\bibitem{Aichelburg:1970dh}
  P.~C.~Aichelburg and R.~U.~Sexl,
  Gen.\ Rel.\ Grav.\  {\bf 2}, 303 (1971).
  
\bibitem{barack_sago}
  L.~Barack and N.~Sago,
  Phys.\ Rev.\ Lett.\  {\bf 102}, 191101 (2009).

\bibitem{Sago:2008id}
  N.~Sago, L.~Barack and S.~L.~Detweiler,
  Phys.\ Rev.\  D {\bf 78}, 124024 (2008).

\bibitem{SFcodes} L.~Barack and N.~Sago, Phys.\ Rev.\ D {\bf 81}, 084021 (2010);  N.~Warburton and L.~Barack,
   Phys.\ Rev.\  D {\bf 81}, 084039 (2010).
  %%CITATION = PHRVA,D81,084039;%%

% L.~Barack, T.~Damour and N.~Sago, Phys.\ Rev.\ D {\bf 82}, 084036 (2010)

 \bibitem{eob} E.~Barausse and A.~Buonanno, Phys.\ Rev.\ D {\bf 81} 084024 (2010).

 \bibitem{favata} M.~Favata, arXiv:1010.2553

\end{thebibliography}
\end{document}